\documentclass[12pt]{iopart}
\usepackage{graphicx}
\usepackage{dcolumn}
\usepackage{bm}
\usepackage{amsfonts}
\usepackage{amssymb}
\usepackage{multirow}
\usepackage{iopams}

\usepackage{epstopdf}
\usepackage[abs]{overpic}

\usepackage{cancel}
\usepackage{color}

\usepackage[dvipsnames]{xcolor}

\newcommand{\blue}{\color{black}}
\newcommand{\red}{\color{black}}

\usepackage{braket}

\definecolor{specialgray}{HTML}{505050}
\definecolor{col10K}{HTML}{FFA000}
\definecolor{col300K}{HTML}{924FA4}
\definecolor{colMu}{HTML}{5278BD}
\definecolor{colMuI}{HTML}{924FA4}
\definecolor{newred}{HTML}{D53E4F}
\definecolor{newblue}{HTML}{5278BD}
\definecolor{newcyan}{HTML}{4EBCB3}
\definecolor{newgreen}{HTML}{5CB14E}
\definecolor{newpurple}{HTML}{924FA4}
\definecolor{newyellow}{HTML}{D1C72E}
\definecolor{neworange}{HTML}{D6923C}

\begin{document}
\title[Computational investigation of superconductivity in a cuprate model]{Doping dependence and multichannel mediators of superconductivity: Calculations for a cuprate model
}
\author{Fabian Schrodi, Alex Aperis\footnote{Present address: 
                     Department of Cell and Molecular Biology, Uppsala University, P.\ O.\ Box 596, SE-75124 Uppsala, Sweden}
 and Peter M. Oppeneer\footnote{Corresponding author}}
\address{Department of Physics and Astronomy, Uppsala University, P.\ O.\ Box 516, SE-75120 Uppsala, Sweden}
\ead{\mailto{fabianschrodi@yahoo.de}, \mailto{alex.aperis@icm.uu.se}, \mailto{peter.oppeneer@physics.uu.se}}
	
\date{\today}

\begin{abstract}
 We study two aspects of the superconductivity in a cuprate model system, its doping dependence and the influence of competing pairing mediators.  We first include  electron-phonon interactions beyond Migdal's approximation and solve self-consistently,
 as a function of doping and for an isotropic electron-phonon coupling, the full-bandwidth, anisotropic vertex-corrected Eliashberg equations under a non-interacting state approximation for the vertex correction. Our results show that such pairing interaction supports the experimentally observed $d_{x^2-y^2}$-wave symmetry of the superconducting gap, but only in a narrow doping interval of the  hole-doped system. Depending on the coupling strength, we obtain realistic values for the gap magnitude and superconducting critical temperature $T_c$ close to optimal  doping, rendering the electron-phonon mechanism an important candidate for mediating superconductivity in this model system.  Second, for a doping near optimal {\blue hole} doping, we study multichannel superconductivity, by including both vertex-corrected electron-phonon interaction and spin and charge fluctuations as pairing mechanisms. We find that both mechanisms cooperate to support an unconventional $d$-wave symmetry of the order parameter, yet the electron-phonon interaction is mainly responsible for the Cooper pairing and high critical temperature $T_c$. Spin fluctuations are found to have a suppressing effect on the gap magnitude and critical temperature due to their repulsive interaction at small coupling wave vectors. 
\end{abstract}

\maketitle

\section{Introduction}

Superconductivity in the cuprate family was discovered more than 35 years ago \cite{Bednorz1986} and has since been a topic of intense investigations.
To this date, the record holder for high-temperature superconductivity with $T_c \approx 130$\,K at ambient pressure is a copper-based compound \cite{Schilling1993}. Despite an immense  effort in theoretical and experimental research spanning several decades, the origin of the high-$T_c$ superconductivity in cuprates remains an unresolved issue. 

There are several disputed aspects in the cuprate superconductivity. One of these is the Cooper pairing mechanism, with the electron-phonon interaction (EPI) and an electronic mechanism, spin fluctuations (SFs), being the most likely candidates.  Cooper pair formation is often associated with electronic mechanisms for various reasons \cite{Sobota2021}. First, many undoped cuprates show antiferromagnetic order and become superconducting only upon sufficient electron or hole doping \cite{Taillefer2010,Rybicki2016}, which naturally gives rise to an association of Cooper pairing due to spin fluctuations. There is also substantial experimental evidence that the proximity to antiferromagnetic order plays a role for superconductivity, because the magnon excitation spectrum in the superconducting state can be interpreted as softened magnons from the undoped parent phase \cite{LeTacon2011,Guarise2014}. 
Second, the observed unconventional $d$-wave symmetry of the superconducting gap \cite{Chen1994,Shen1995,Kirtley1995,Shi2008} 
can be explained because the SF interaction follows approximately the antiferromagnetic wave vector, as has been pointed out in theoretical investigations,  see Refs.\,\cite{Schmalian1996,Moriya2006,Scalapino2012} and references therein.

Although ARPES measurements have shown that phonons heavily influence the electron dynamics \cite{Lanzara2001,Yang2019},  and other relevant signatures of EPI, such as the isotope effect have been found as well \cite{Iwasawa2008,Verga2003,Berthod2017,Shen2002,Devereaux2004}, electron-phonon coupling is often discarded as mediator for high-temperature superconductivity in the cuprates.  An argument against EPI is the fact that an isotropic, attractive coupling cannot lead to a $d$-wave order parameter within standard Bardeen-Cooper-Schrieffer (BCS) or Eliashberg theories. It has been argued that strongly peaked forward scattering could lead to a sign-changing gap \cite{Kulic2006,Varelogiannis2007,Weger1996,Weger1997}, but this has not changed the view that EPI is not a likely contender for the pairing mechanism in cuprates.

A further intriguing aspect of high-temperature cuprate superconductivity is the origin of its strong doping dependence \cite{Proust2019}.  Superconductivity appears at a certain hole doping away from the antiferromagnetic phase and emerges as a dome-shaped region in the {\blue hole-doped} phase diagram, where there exists an optimal doping associated with a maximum critical temperature \cite{Taillefer2010,Rybicki2016}. 

It needs to be stressed that the above conclusions regarding the EPI rely on \textit{adiabatic} Eliashberg theory \cite{Eliashberg1960,Marsiglio2020}, which is based on Migdal's approximation for the EPI \cite{Migdal1958}, stating that only first-order electron-phonon scattering events are important to describe the interacting state in a superconductor. This holds  when $\alpha$, the ratio of the phonon energy scale over that of the electrons, is much smaller than one, and, hence, vertex corrections can be neglected.  
For cuprates this condition is however not fulfilled as $\alpha$ often reaches values of $\alpha \gtrsim0.1$, while the precise value depends on the doping level and the specific compound. 

To shed light on aspects of superconductivity, self-consistent calculations of the superconducting state are desirable, as these can provide microscopic understanding \cite{Boeri2022}. In this way, self-consistent solutions to the adiabatic Eliashberg equations have elucidated the two-band superconductivity in MgB$_2$
\cite{Choi2002,Choi2006}. Superconductivity in hydrogen and super-hydrides under high pressure has been explained using density-functional theory for superconductors (SCDFT) \cite{Cudazzo2008,Flores2016,Flores2020} as well as adiabatic Eliashberg theory \cite{Verma2021,Lucrezi2024}.

In the current work we follow this route, but we employ non-adiabatic Eliashberg theory, since it has been shown recently that including vertex corrections beyond Migdal's approximation can lead to negative contributions to the electron-phonon coupling \cite{Schrodi2020_2}, leading to an effectively attractive (repulsive) coupling at small (large) wave vectors, thereby overall supporting a $d$-wave symmetric superconducting gap \cite{Schrodi2021}. Not only does the inclusion of second-order scattering events increase the coupling strength in the $d$-wave channel \cite{Ishihara2004,Ishihara2006}, it has also been shown that self-consistent non-adiabatic model calculations assuming an isotropic EPI can lead to the experimentally observed unconventional symmetry of the order parameter in slightly hole-doped cuprates \cite{Schrodi2021,Hague2006}. Apart from the gap symmetry, the gap magnitude and critical temperature $T_c$ are similarly found to be in a realistic regime when employing vertex-corrected Eliashberg theory \cite{Schrodi2021}. 

In the following, we solve the vertex-corrected, {full-bandwidth} Eliashberg equations for isotropic electron-phonon coupling in a cuprate model system, as function of doping, temperature, and scattering strength. As it is not unlikely that both phonons and spin fluctuations contribute in mediating the record critical temperatures, we employ in a second step a multichannel Eliashberg formalism, i.e., we consider both  EPI and SFs \cite{Schrodi2020_3}. For the electronic part of the interaction we employ a single-orbital flucuation-exchange (FLEX) formalism \cite{Bickers1989}, while the EPI is treated beyond Migdal's approximation \cite{Migdal1958}, taking into account all first and second order electron-phonon scattering processes \cite{Schrodi2020_2}. With the resulting vertex-corrected Eliashberg theory we are capable of studying the cooperative and competitive effects of both mediators of superconductivity. 

Our calculations show that vertex corrections support a $d$-wave symmetric superconducting gap only in a restricted doping regime, on the hole-doped side, bearing strong similarities to the dome-like superconducting phase commonly observed in {\blue hole-doped} cuprate systems. We further identify the electron-phonon scattering strength as important factor when it comes to finding a self-consistent $d$-wave solution from electron-phonon interactions only. This is due to the interplay between the first- and second-order contributions to the overall interaction, which is required to change sign in momentum space to support an unconventional gap. 

Since our calculations show that vertex-corrected EPI can provide the $d$-wave gap symmetry and a dome-like superconducting phase, explicit calculations with both mediators of superconductivity on equal footing are required to determine how their roles are distributed.
Our self-consistent, anisotropic full-bandwith calculations reveal that SFs and charge fluctuations (CFs) alone do not lead to a stable superconducting solution except at very low temperatures.  This stems from the small wave-vector contributions to the SF interaction, which support a different gap symmetry than the dominant coupling around the nesting wave vector, an effect that has been observed, too, in Fe-based superconductors \cite{Yamase2020,Schrodi2020_3}. When both interactions are included we obtain typical $d$-wave solutions with sizable $T_c$, where in general an increase in the electron-phonon coupling strength increases $T_c$, while stronger influence of the electronic mechanism leads to a reduction of the critical temperatures, indicating that EPI could be the prime superconductivity mediator. Our work is a step towards realistic multichannel superconductivity calculations for high-$T_c$ materials, in which the interplay of EPI with purely electronic mechanisms can be thoroughly analyzed in one framework.

\section{Theoretical Methodology}

\subsection{The electron-phonon interaction}\label{epi}

We start our theoretical  framework to describe electron-phonon mediated superconductivity similar to Ref.\,\cite{Schrodi2020_2}, with the Hamiltonian 
\begin{equation}
\hat{{\cal H}}^{(\rm ep)} = \sum_{\mathbf{k}} \xi_{\mathbf{k}} \Psi_{\mathbf{k}}^{\dagger} \hat{\rho}_3 \Psi_{\mathbf{k}}^{ } +  \sum_{\mathbf{q}} 
\Omega\Big( b_{\mathbf{q}}^{\dagger}b_{\mathbf{q}}^{}  + \frac{1}{2} \Big) 
+ \sum_{\mathbf{k},\mathbf{q}} g_{\mathbf{q}} u_{\mathbf{q}} \Psi_{\mathbf{k}-\mathbf{q}}^{\dagger} \hat{\rho}_3 \Psi_{\mathbf{k}}^{} . \label{hamiltonian}
\end{equation}
The effects due to direct Coulomb repulsion are assumed to be included in our square lattice tight-binding description of the electron energies $\xi_{\mathbf{k}}$, and its influence on the interaction is neglected for simplicity. Electrons are created and annihilated via $c^{\dagger}_{\mathbf{k},\sigma}$ and $c_{\mathbf{k},\sigma}^{}$, with $\sigma$ a spin index, which are contained in Nambu \cite{Nambu1960} spinors $\Psi^{\dagger}_{\mathbf{k}}=(c^{\dagger}_{\mathbf{k},\uparrow},c_{-\mathbf{k},\downarrow})$ spanning a $2\times2$ pseudovector space, and the  $\hat{\rho}_i$ are the Pauli matrices. We model the phonon spectrum via a single Einstein {phonon energy}
$\Omega$ {\red (using $\hbar =1$)}, with associated creation and annihilation operators $b^{\dagger}_{\mathbf{q}}$ and $b_{\mathbf{q}}$. The EPI in Eq.\,(\ref{hamiltonian}) is given by the phonon displacement $u_{\mathbf{q}}=(b^{\dagger}_{\mathbf{q}}+b_{-\mathbf{q}})$ and scattering matrix elements $g_{\mathbf{q}}$. The latter are assumed to be isotropic in this work, i.e., $g_{\mathbf{q}}=g_0$. 

The interacting state of the system is characterized by the electronic self-energy $\hat{\Sigma}_k$ and electron Green's function $\hat{G}_k$, where we use the four-momentum notation $k=(\mathbf{k},i\omega_m)$, $q=(\mathbf{q},iq_l)$, with fermionic Matsubara frequencies $\omega_m=\pi T(2m+1)$  and bosonic Matsubara  frequencies $q_l=2\pi Tl$ {\red ($m, \, l \in \mathbb Z $)}, {\blue at temperature $T$}. As the Nambu space is spanned by the Pauli matrices $\hat{\rho}_i$,  the electron self-energy can be written as a matrix decomposition
\begin{equation}
\hat{\Sigma}_k = i\omega_k(1-Z_k)\hat{\rho}_0 + \chi_k\hat{\rho}_3 + \phi_k\hat{\rho}_1 . \label{sigmaDef}
\end{equation}
Here, $Z_k$ is the mass renormalization, $\chi_k$ the chemical potential renormalization, and $\phi_k$ the superconductivity order parameter that we aim to computed self-consistently. 
Note that the decomposition (\ref{sigmaDef}) is generally valid for different interactions contributing to the self-energy.

In the non-interacting state the electron Green's function has the well-known form $\hat{G}_k^{(0)}=[i\omega_k\hat{\rho}_0-\xi_k\hat{\rho}_3]^{-1}$, which, together with Eq.\,(\ref{sigmaDef}), determines the functional form of the Dyson equation $\hat{G}_k=\hat{G}^{(0)}_k + \hat{G}^{(0)}_k\hat{\Sigma}_k\hat{G}_k$ for the fully interacting Green's function.  By using the definitions $\gamma_k^{(Z)}=\omega_kZ_k/\Theta_k$, $\gamma_k^{(\chi)}=(\xi_k+\chi_k)/\Theta_k$ and $\gamma_k^{(\phi)}=\phi_k/\Theta_k$, with
\begin{equation}
\Theta_k = \big(i\omega_k Z_k\big)^2 - \big(\xi_k+\chi_k\big)^2 - \phi_k^2 ,
\end{equation}
the above expressions lead to
\begin{equation}
\hat{G}_k = i\gamma_k^{(Z)}\hat{\rho}_0 + \gamma_k^{(\chi)}\hat{\rho}_3 + \gamma_k^{(\phi)} \hat{\rho}_1 .
\end{equation}

\begin{figure*}[ht!]
	\centering
	\includegraphics[width=0.7\columnwidth]{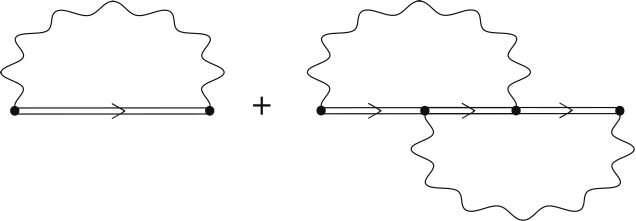}
	\caption{\blue The first- and second-order Feynman diagrams of the electron self-energy $\hat{\Sigma}^{(\rm ep)}_k$ due to the electron-phonon interaction included in the here-used vertex-corrected Eliashberg formalism. Straight double lines are full electron Green’s functions and wavy lines are bare phonon propagators.} \label{diagrams}
\end{figure*}

As mentioned in the Introduction, we take the electron-phonon interaction into account beyond Migdal's approximation. {\blue Specifically, this is the lowest order electron-phonon interaction customarily used in Migdal's approximation and its vertex correction. The  corresponding Feynman diagrams are shown in Fig.\ \ref{diagrams}.}
The electron self-energy including these first two Feynman diagrams for electron-phonon scattering  can be written as {\blue \cite{Schrodi2020_2}}
\begin{equation}
\hat{\Sigma}^{(\rm ep)}_k = T\sum_{k_1} V^{(\mathrm{ep})}_{k-k_1} \hat{\rho}_3 \hat{G}_{k_1} \hat{\rho}_3 \big(1+g_0^2\hat{\Gamma}_{k,k_1} \big) ,
\label{SigmaEPI}
\end{equation}
where the fully self-consistent vertex function is given by
\begin{equation}
\hat{\Gamma}_{k,k_1} = \frac{T}{g_0^2} \sum_{k_2} V^{(\mathrm{ep})}_{k_1-k_2} \hat{G}_{k_2} \hat{\rho}_3 \hat{G}_{k_2-k_1+k}\hat{\rho}_3 . \label{vertex}
\end{equation}
Here, 
we denote $V^{(\mathrm{ep})}_{k-k_1}
=2g_0^2\Omega/(\Omega^2+q^2_{m-m_1})$ as the electron-phonon interaction kernel.

To reduce the numerical workload we introduce here a non-interacting state approximation for the vertex, or more specifically, for the  electron Green's functions $\hat{G}_{k}$ in Eq.\,(\ref{vertex}). The accuracy of this approximation for cuprate systems will be analyzed in Sec.\ \ref{doping-approx}. Specifically, in this approximation the vertex simplifies to
\begin{equation}
\hat{\Gamma}_{k,k_1} \simeq \hat{\Gamma}_{k,k_1}^{(0)} = \frac{T}{g_0^2} \sum_{k_2} V_{k_1-k_2}^{(\mathrm{ep})} \hat{G}_{k_2}^{(0)} \hat{\rho}_3 \hat{G}_{k_2-k_1+k}^{(0)}\hat{\rho}_3 , \label{nonint}
\end{equation}
so that with definitions $\gamma_k^{(\xi)}=\xi_k/\theta_k$, $\gamma_k^{(\omega)}=\omega_k/\theta_k$ and $\theta_k = -\omega_k^2 - \xi_k^2$ we obtain
\begin{eqnarray}
\hat{\Gamma}_{k,k_1}^{(0)} =& \frac{T}{g_0^2} \sum_{k_2} V^{(\mathrm{ep})}_{k_1-k_2} \big[ \big(\gamma_{k_2}^{(\xi)}\gamma_{k_2-k_1+k}^{(\xi)} - \gamma_{k_2}^{(\omega)}\gamma_{k_2-k_1+k}^{(\omega)} \big) \hat{\rho}_0  \nonumber \\
&~~~~~~+ \big(\gamma_{k_2}^{(\xi)}\gamma_{k_2-k_1+k}^{(\omega)}  + \gamma_{k_2}^{(\omega)}\gamma_{k_2-k_1+k}^{(\xi)}\big) i\hat{\rho}_3  \big] . \label{gamma0}
\end{eqnarray}
Note, that $\hat{\Gamma}_{k,k_1}^{(0)}$ can be pre-computed and hence, this approach offers a significant numerical advantage.

The here-made truncation of the infinite series of Feynman diagrams for the electron-phonon interaction after the second-order scattering terms is valid for weakly non-adiabatic systems {\blue and when we are in the non-strong coupling regime}. This is when the non-adiabaticity ratio \cite{Migdal1958}, $ \alpha =  \Omega/\varepsilon_{\rm F}$, where $\Omega$ is the energy scale of the phonons and $\varepsilon_F$ that of the electrons, is small,
typically of the order $\alpha = 0.1 - 0.2$ \cite{Schrodi2020_2,Schrodi2021}.

From the above definitions and relations we can derive a set of three self-consistent, {non-adiabatic} Eliashberg \cite{Eliashberg1960} equations. These will be specified further below in Sec.\  \ref{eliash}, but first spin and charge fluctuations will be included in the methodology.

\subsection{Spin and charge fluctuations}\label{scf}

To take  SFs and CFs into account we can write the electronic self-energy as
\begin{equation}
\hat{\Sigma}_k = \hat{\Sigma}_k^{(\mathrm{ep})} + \hat{\Sigma}_k^{(\mathrm{s})} + \hat{\Sigma}_k^{(\mathrm{c})} , \label{sigmaFull}
\end{equation}
with labels $(\mathrm{s)}$ and $(\mathrm{c})$ for SFs, and CFs, respectively. 
Assuming spin-singlet Cooper pairs, we treat here the purely electronic parts of the electron self-energy similarly to Ref.\,\cite{Lenck1994}, i.e., within a FLEX formalism \cite{Bickers1989}. The self-energy contributions to SFs and CFs are calculated via
\begin{eqnarray}
\hat{\Sigma}_k^{(\rm s)} &= T\sum_{k_1} V^{(\rm s)}_{k-k_1} \hat{\rho}_0 \hat{G}_{k_1} \hat{\rho}_0 ,\\
\hat{\Sigma}_k^{(\rm c)} &= T\sum_{k_1} V^{(\rm c)}_{k-k_1} \hat{\rho}_3 \hat{G}_{k_1} \hat{\rho}_3 ,
\end{eqnarray}
where the interactions are defined in terms of the single-orbital, on-site 
$U$ parameter:
\begin{eqnarray}
V^{(\rm s)}_{q} = \frac{3}{2} U^2 X^{(\rm s)}_q  , \\
V^{(\rm c)}_{q} = \frac{1}{2} U^2 X^{(\rm c)}_q  .
\end{eqnarray}
{\blue For details of the derivation we refer to Refs.\ \cite{Wermbter1991,Lenck1992,Kubo2007}  and references therein.} Note that this $U$ parameter is not the same as the on-site electron Coulomb repulsion that is included already in the  electron energies $\xi_{\mathbf{k}}$.

The electron spin and charge susceptibilities in the interacting state are calculated from the random-phase-approximation-like expressions
\begin{eqnarray}
X^{(\rm s)}_q &= \frac{X^{(\rm s,0)}_q}{1-UX^{(\rm s,0)}_q} , \\
X^{(\rm c)}_q &= \frac{X^{(\rm c,0)}_q}{1+UX^{(\rm c,0)}_q} ,
\end{eqnarray}
while the non-renormalized functions are given in terms of the system's Green's function,
\begin{eqnarray}
X^{(\rm s,0)}_q &= -\frac{T}{2}\sum_k \mathrm{Tr} \big\{ \hat{G}_k \hat{\rho}_0 \hat{G}_{k+q} \hat{\rho}_0 \big\} , \label{xs0}\\
X^{(\rm c,0)}_q &= -\frac{T}{2}\sum_k \mathrm{Tr} \big\{ \hat{G}_k \hat{\rho}_3 \hat{G}_{k+q} \hat{\rho}_3 \big\} . \label{xc0}
\end{eqnarray}
In the above equations we use the fully interacting Green's function, hence the SF and CF spectra are iteratively updated making our calculations fully self-consistent. 
{\blue The $U$ parameter can be varied to approach the spin-fluctuation instability, marked by the condition ${1-UX^{(\rm s,0)}_q} \rightarrow 0$.}

\subsection{Eliashberg equations}\label{eliash}

By using the standard procedure we derive the Eliashberg equations from the two expressions for the self-energy Eqs.\,(\ref{sigmaDef}) and (\ref{sigmaFull}), and find
\begin{eqnarray}
Z_k &= 1 - \frac{T}{\omega_k} \! \sum_{k_1} \Big[ \big( V^{(+)}_{k-k_1} +  V^{(\mathrm{ep},1)}_{k,k_1} \big) \gamma_{k_1}^{(Z)} +   {V}^{(\mathrm{ep},2)}_{k,k_1}\gamma_{k_1}^{(\chi)} \Big] , \label{z}\\
\chi_k &= T \sum_{k_1} \Big[\big( V^{(+)}_{k-k_1} + V^{(\mathrm{ep},1)}_{k,k_1} \big) \gamma_{k_1}^{(\chi)} - {V}^{(\mathrm{ep},2)}_{k,k_1}\gamma_{k_1}^{(Z)}  \Big]  , \label{chi}\\
\phi_k &= -T \sum_{k_1} \big(  V^{(-)}_{k-k_1}  + V^{(\mathrm{ep},1)}_{k,k_1} \big) \gamma_{k_1}^{(\phi)}  . \label{phi}
\end{eqnarray}
Herein, the first-order interaction kernels containing EPI, SFs, and CFs are defined as
\begin{equation}
V_{k-k_1}^{(\pm)} = V_{k-k_1}^{(\mathrm{ep})} + V_{k-k_1}^{(\mathrm{c})} \pm V_{k-k_1}^{(\mathrm{s})} ,
\end{equation}
while the second-order EPI kernels  due to the vertex correction within the non-interacting state approximation are given by
\begin{eqnarray}
V^{(\mathrm{ep},1)}_{k,k_1} = T V^{(\mathrm{ep})}_{k-k_1} \sum_{k_2} V^{(\mathrm{ep})}_{k_1-k_2}  \big( \gamma_{k_2}^{(\xi)}\gamma_{k_2-k_1+k}^{(\xi)} -\gamma_{k_2}^{(\omega)}\gamma_{k_2-k_1+k}^{(\omega)} \big) , \label{kern1}\\ 
{V}^{(\mathrm{ep},2)}_{k,k_1} = T V^{(\mathrm{ep})}_{k-k_1} \sum_{k_2} V^{(\mathrm{ep})}_{k_1-k_2} 
 \big( \gamma_{k_2}^{(\xi)}\gamma_{k_2-k_1+k}^{(\omega)}    + \gamma_{k_2}^{(\omega)}\gamma_{k_2-k_1+k}^{(\xi)}  \big) . \label{kern2}
\end{eqnarray}
When solving Eqs.\,(\ref{z})--(\ref{phi}) self-consistently we can either include SFs and CFs, or set these to zero {\blue ($U=0$)} and only consider EPI.  As for the latter, we already mention here that we include both $V^{(\mathrm{ep},1)}_{k,k_1}$ and ${V}^{(\mathrm{ep},2)}_{k,k_1}$ in our numerical calculations, but, as shown in more detail below, contributions from ${V}^{(\mathrm{ep},2)}_{k,k_1}$ are generally an order of magnitude smaller than those of $V^{(\mathrm{ep},1)}_{k,k_1}$. {\blue Moreover, the superconducting order parameter $\phi_k$ depends primarily on $V^{(\mathrm{ep},1)}_{k,k_1}$, see Eq.\ (\ref{phi}).} For this reason we will not analyze the former coupling terms in detail in the following sections. 

The here-presented formalism is a step forward to combining SFs and CFs with phonons in vertex-corrected Eliashberg theory, yet it is appropriate to briefly recall  the assumptions made in the formalism. The SFs and CFs are included on the level of the FLEX approach, while EPI  is included up to all second-order processes  within the non-interacting state approximation for two of the three electron Green's functions appearing in the vertex correction of Eq.\ (\ref{SigmaEPI}). 
 The Cooper-pair breaking Coulomb interaction is assumed to be already included in 
the electronic energy dispersions {and we omit its effect on the off-diagonal self-energies for simplicity}. It should further be noted that we neglect the phonon self-energy here, which might be important for high-temperature superconductors \cite{Zeyher1990}. {Also,} both mediators of superconductivity are {included but} treated independently of each other. 
These approximations are made for computational accessibility, so as to obtain quantitative results. 

\begin{figure}[t!]
	\centering
	\includegraphics[width=0.45\columnwidth]{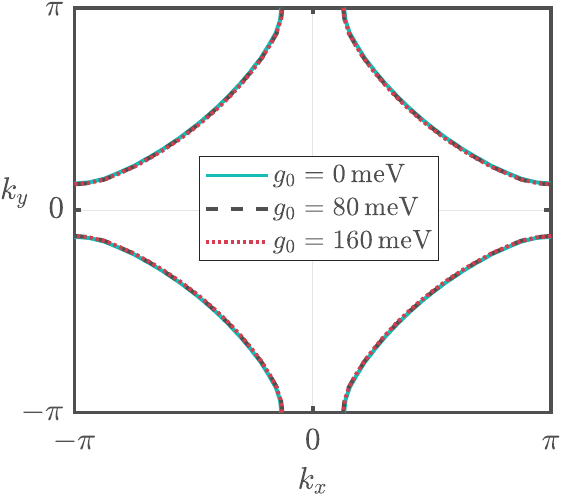}
	\caption{Fermi surface of a model cuprate calculated for $T>T_c$ and $\mu=-0.07\,\mathrm{eV}$ for {various electron-phonon} scattering strengths $g_0$ as listed in the legend.}	\label{fs}
\end{figure}

The numerical calculations in this work have been carried out with the Uppsala Superconductivity Code (\textsc{uppsc}) {\cite{UppSC,Aperis2015,Schrodi2019,Schrodi2020_2,Schrodi2021_2}. For solving Eliashberg-type coupled equations in principle infinite Matsubara sums have to be carried out, which poses a computational difficulty.  To achieve the required full numerical convergence we have employed specifically designed numerical schemes that considerably improve the numerical accuracy \cite{Schrodi2019}. 

\section{Results}
\label{Results}

\subsection{Electronic structure model}

Our aim is to describe the cuprate family of superconductors, which have a quasi two-dimensional structure. {\blue In this work, we focus on the typical class of hole-doped cuprates.} 
 The electron energies are given by a next-nearest neighbor tight-binding model
\begin{equation}
\xi_{\mathbf{k}} = t\big[\cos(k_x) + \cos(k_y)\big] + t'\cos(k_x)\cos(k_y) - \mu = \xi_{\mathbf{k}}^0 - \mu ,
\end{equation}
where the hopping energies are chosen as $t=-0.25\,\mathrm{eV}$ and $t'=0.1\,\mathrm{eV}$ and $\mu$ is {\blue a parameter determining the initial} chemical potential. 
To simulate the effect of doping, we {\blue will vary $\mu$ and treat it as a free parameter. We analyze below that this is a suitable approximation.} 
We further choose the  Einstein phonon frequency $ \Omega$ as $50\,\mathrm{meV}$, which is in accordance with experiments performed on representative examples of the cuprate family \cite{Lanzara2001,Sobota2021}. We mention that we have repeated several of the calculations presented below for $\Omega=70\,\mathrm{meV}$ and obtained qualitatively similar results. 
The other two parameters in our simulations are the electron-phonon scattering strength $g_0$ and the $U$ parameter. For one-dimensional cuprates electron-phonon interaction strengths in the range of 100 to 200 meV have been deduced from experiments \cite{Wang2021}. Varying the  $U$ parameter can be used to strengthen the influence of SFs and CFs on the superconductivity \cite{Schrodi2020_4}. In the following we will first evaluate the self-consistent phonon-mediated superconductivity and then simulate the influence of doping on the superconductivity by varying the 
{\blue parameter $\mu$} for  $g_0$ values between 100 and 200 meV, and $U=0$. After that, we will investigate the influence of SFs and CFs and EPI in vertex-corrected self-consistent Eliashberg theory.

\subsection{Influence of electron-phonon coupling}

To start with, we consider only the EPI and solve self-consistently the full-bandwidth anisotropic Eliashberg equations with vertex corrections.
We choose  $g_0=170\,\mathrm{meV}$ {\blue and $\mu=-0.07$ eV,} and compute the EPI kernel $V^{(\mathrm{ep})}_{\mathbf{q},l}+V^{(\mathrm{ep,1})}_{\mathbf{q},m,m'}$, which enters each of Eqs.\,(\ref{z})-(\ref{phi}). In Fig.\,\ref{gapexample}(a) we show this interaction kernel in the 2D Brillouin zone (BZ) for  $m=m'=0$  and $T=80\,\mathrm{K}$.  In accordance with Ref.\,\cite{Schrodi2021}, this function is repulsive around the nesting wave vector $\mathbf{q}\sim(-\pi,\pi)$, i.e., the M point, and attractive in most parts of the remaining BZ. 
{\red As shown in Fig.\  \ref{gapexample}(b), the strongly repulsive nature at the M point stems fully from $V^{(\mathrm{ep,1})}_{\mathbf{q},m=0,m'=0}$. This kernel is quite small in most of the BZ, but it peaks at the M point reaching a value of $-1970$ meV.  Conversely, $V^{(\mathrm{ep})}_{\mathbf{q},l=0} = 1156$ meV is  large and attractive throughout the whole BZ.} Such characteristics are favorable for a $d$-wave symmetry of the superconducting gap. In Fig.\,\ref{gapexample}{\red (c)} we show $V^{(\mathrm{ep,2})}_{\mathbf{q},m=0,m'=0}$, which is negative in most parts of the BZ and significantly smaller in magnitude than the interaction in panel (a). It further deserves mentioning that $V^{(\mathrm{ep,2})}_{\mathbf{q},m,m'}$ enters only in Eqs.\,(\ref{z}) and (\ref{chi}), and therefore it should not have a big impact on the superconducting properties.

\begin{figure*}[bht!]
	\hspace{0.8cm}
	\includegraphics[width=1\textwidth,bb= 1 30 600 500]{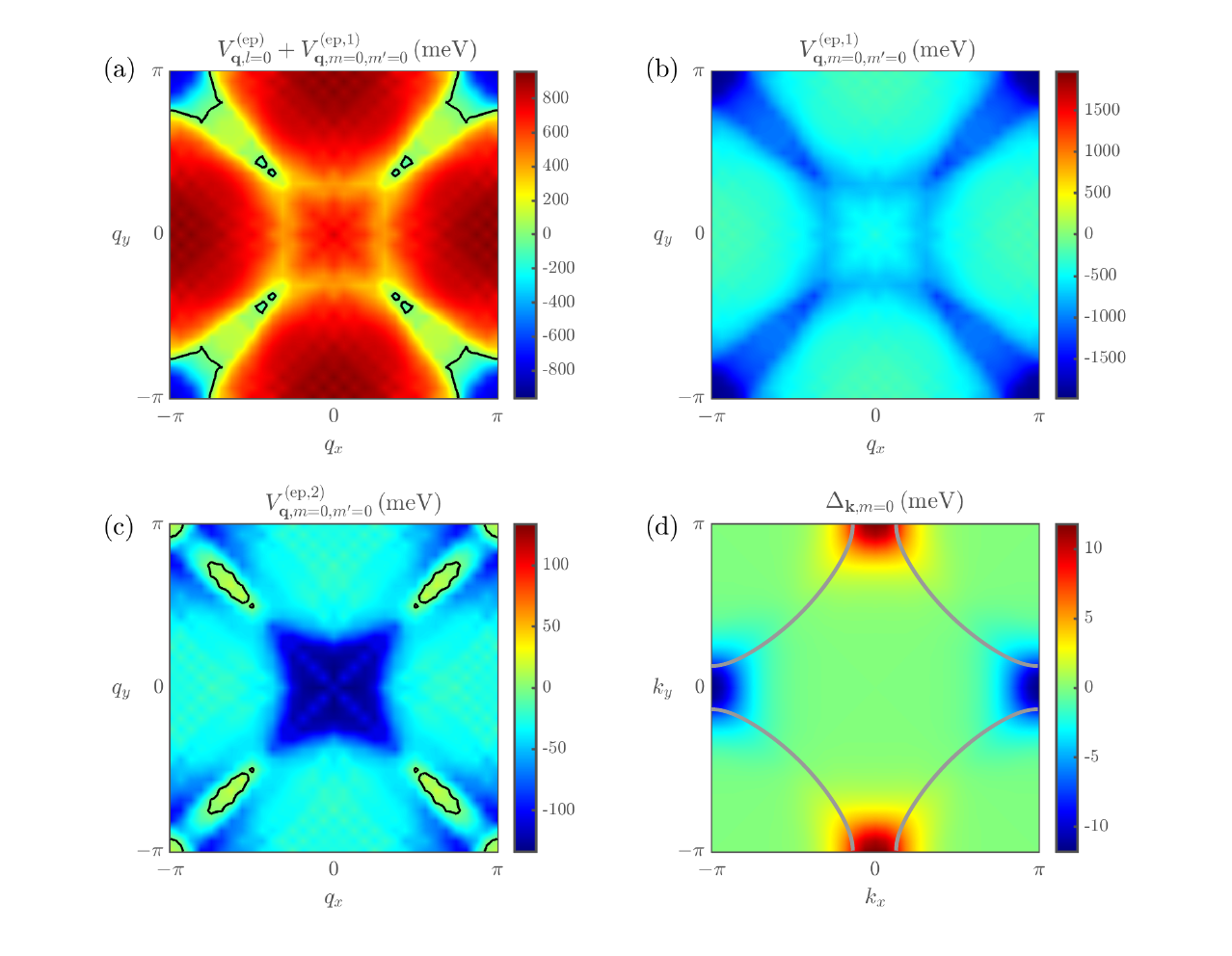}
	\caption{(a) Electron-phonon interaction kernels $V^{(\mathrm{ep})}_{\mathbf{q},m-m'}+V^{(\mathrm{ep},1)}_{\mathbf{q},m,m'}$ at $m=m'=0$, calculated for $g_0=170\,\mathrm{meV}$, {\blue $U= 0$ eV}, and $T=80\,\mathrm{K}$. (b) {\red Same as (a), but for the vertex correction to the electron-phonon kernel, $V^{(\mathrm{ep},1)}_{\mathbf{q},m,m'}$. (c)} Same as (a), but for the {second} correction to the 
	electron-phonon kernel, $V^{(\mathrm{ep},2)}_{\mathbf{q},m,m'}$. Solid black lines indicate zero crossings of the kernel functions. (c) Self-consistent zero-frequency component of the superconducting gap, having $d_{x^2-y^2}$-symmetry, obtained for $g_0=170\,\mathrm{meV}$, $T=80\,\mathrm{K}$ and $U=0\,\mathrm{eV}$. The cuprate Fermi surface is shown in gray color.}	\label{gapexample}
\end{figure*}

In Fig.\,\ref{gapexample}{\red (d)} the cuprate Fermi surface is shown through the gray lines and the self-consistently calculated zero-frequency gap function $\Delta_{\mathbf{k},m=0}$ is given by the colors. 
When solving the Eliashberg equations self-consistently for this set of parameters we find the experimentally observed $d_{x^2-y^2}$ symmetry of the  gap function $\Delta_{\mathbf{k},m=0}=\phi_{\mathbf{k},m=0}/Z_{\mathbf{k},m=0}$. This is the only symmetry of the superconducting order parameter that we found here. Therefore, when discussing superconductivity for this system with any other parameter settings below, everything is to be understood in terms of a $d$-wave symmetric gap function. The chosen value $g_0=170\,\mathrm{meV}$ corresponds to an input electron-phonon coupling strength of $\lambda_0=2$. The \textit{effective} interaction strength $\lambda_m$ due to electron-phonon interaction can be calculated above $T_c$ \cite{Schrodi2021}, via
\begin{equation}
\langle Z_{\mathbf{k},m=0}\rangle_{\mathbf{k_F}}|_{T>T_c} = 1+\lambda_m - \lambda_m^2 \frac{\pi^2\Omega N_0}{4}\frac{3\sinh{(\frac{\Omega}{T})}-\frac{\Omega}{T}}{\cosh{(\frac{\Omega}{T})-1}} .
\end{equation}
At $T=120\,\mathrm{K}$ (above $T_c$)  we find $\lambda_m=1.5$, which is a significant reduction compared to the input coupling. 
Note, that even more realistic values for the electron-phonon coupling strength {\red ($\lambda < 1$)}  as compared to experiments \cite{Lanzara2001} are {obtained} when the full-temperature dependent vertex-corrected Eliashberg equations are solved \cite{Schrodi2021}. 

\subsection{Approximations}
\label{doping-approx}

Next, we want to study the influence of doping dependence on the superconductivity. To make the computational effort manageable, we make two approximations.
First, we introduce the non-interacting state approximation for the electron-phonon vertex, by replacing the electron Green's functions $\hat{G}_k$ by $\hat{G}_k^{(0)}$ in Eq.\ (6). 
Second, we treat the {\blue bandfilling 
using $\mu$ as a free parameter, that is, we do not compute self-consistently the electron filling of the bands that is related to $\chi_k$ and the corresponding self-consistently renormalized chemical potential $\mu^{r}$. The latter quantity is temperature dependent and slightly different from the initial $\mu$ \cite{Marel1990,Rietveld1990,Schrodi2018,Chavez2022}}.

\begin{figure}[t!]
	\centering
	\includegraphics[width=0.7\columnwidth]{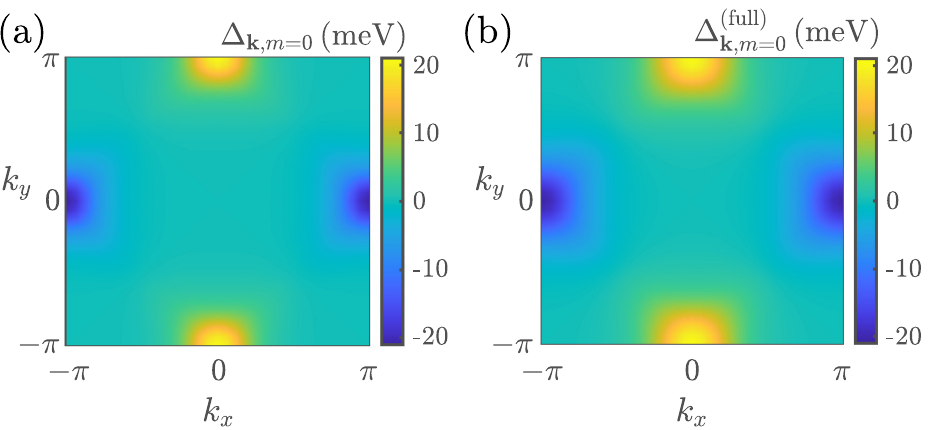}
	\caption{Self-consistently calculated superconducting gap function at zero frequency, using $\Omega=50\,\mathrm{meV}$, $g_0=175\,\mathrm{meV}$, {\blue $U = 0$ eV,} and $T=80\,\mathrm{K}$. (a) The gap function computed using the non-interacting state approximation for the vertex. (b) The complete solution computed without any approximation.}
	\label{benchmark}
\end{figure}

To illustrate the reliability of the first approximation, we perform calculations with the full formalism as well as with the non-interaction state approximation. We use a phonon frequency $\Omega=50\,\mathrm{meV}$, set $g_0=175\,\mathrm{meV}$ {\blue and $\mu= -0.07$ eV}, and we choose the relatively high temperature $T=80\,\mathrm{K}$, to reduce the required number of Matsubara frequencies. In Fig.\,\ref{benchmark} we show a comparison of the obtained results for $\Delta_{\mathbf{k},m=0}$. The superconducting gap function computed with the non-interacting formalism is shown in  Fig.\,\ref{benchmark}(a) and that using the full formalism, labeled `(full)', is shown in  Fig.\,\ref{benchmark}(b).  We observe the same $d$-wave symmetry of the superconductivity order parameter, with remarkably similar extremum $\max_{\mathbf{k}}\Delta_{\mathbf{k},m=0}^{(\mathrm{full})}=20.5\,\mathrm{meV}$, to be compared with $\max_{\mathbf{k}}\Delta_{\mathbf{k},m=0}=21.1\,\mathrm{meV}$. The same holds for the electron mass renormalization, which now takes on a maximum value of $\max_{\mathbf{k}}Z^{(\mathrm{full})}_{\mathbf{k},m=0}=1.70$ as compared  to $\max_{\mathbf{k}}Z_{\mathbf{k},m=0}=1.72$ in the non-interacting state approximation. From these outcomes it is evident that the non-interacting state approximation to the vertex is reliable for the cuprate model system used in the current work.

The second approximation is the treatment of the doping.
In Eliashberg calculations involving doping, it is common practice  {\blue to employ the expression for the electron filling,}
\begin{equation}
n = 1 - 2T\sum_{k} \frac{\xi_k^0 -{\mu^{r}} + \chi_k}{\Theta_k}, \label{filling}
\end{equation}
{\blue see Refs.\ \cite{Lenck1994,Schrodi2018} for details.} {\blue For a chosen electron filling, this equation can be inverted to obtain the corresponding renormalized $\mu^{r}$.}  However, due to the immense computational costs of such calculations \cite{Schrodi2018} it is rather impractical to use Eq.\,(\ref{filling}).} Instead, we {\blue use here the initial chemical potential $\mu$ as free parameter.} 
To show that this procedure is a good approximation, we consider $\mu=-0.07\,\mathrm{eV}$ for the moment. The Fermi surface of the non-interacting system is shown in Fig.\,\ref{fs} as solid cyan lines. For two electron-phonon scattering strengths $g_0=80\,\mathrm{meV}$ and $g_0=160\,\mathrm{meV}$ we solve the Eliashberg equations in the normal state, i.e., for $T>T_c$. The self-consistent results {\blue then} define the renormalized Fermi surface via the condition
\begin{equation}
(\xi_{\mathbf{k}}^0 -\mu^{r} + \chi_{\mathbf{k},m=0})/Z_{\mathbf{k},m=0} = 0 .
\end{equation}
 The results for the Fermi surface in the interacting state for both choices of $g_0$, shown in Fig.\,\ref{fs}, fall on top of the bare Fermi surface. 
Other {\blue choices of the initial chemical potential $\mu$} lead to similar results. Therefore, we can assume that {\blue the electron filling behaves approximately linear with respect to the parameter $\mu$},
if (i) the scattering strength lies in the regime tested in Fig.\,\ref{fs}, i.e., if we are not in the strong coupling limit, and (ii) we don't consider extreme doping levels \cite{Schrodi2020}.

\subsection{Calculated doping dependent superconductivity}
\label{Doping-dependence}

Having verified the reliability of our approximations, we study the influence of doping dependence on the superconductivity,  by  solving the {vertex-corrected} Eliashberg equations in the superconducting state for different values of $\mu$ and $g_0$. Due to the non-interacting state approximation for the vertex function we can pre-calculate the interaction kernels of Eqs.\,(\ref{kern1}) and (\ref{kern2}), resulting in a significant reduction of the computational costs. 
{\blue When studying the influence of doping, we emphasize that our focus is on the unconventional superconductivity appearing in the hole-doped region of the typical cuprate phase diagram, {\red i.e., where our electronic structure model gives a valid description.} The superconductivity emerging in the strongly electron-doped region of the cuprate phase diagram \cite{Taillefer2010,Rybicki2016} is not considered here.}

\begin{figure}[t!]
	\centering
	\includegraphics[width=0.65\columnwidth]{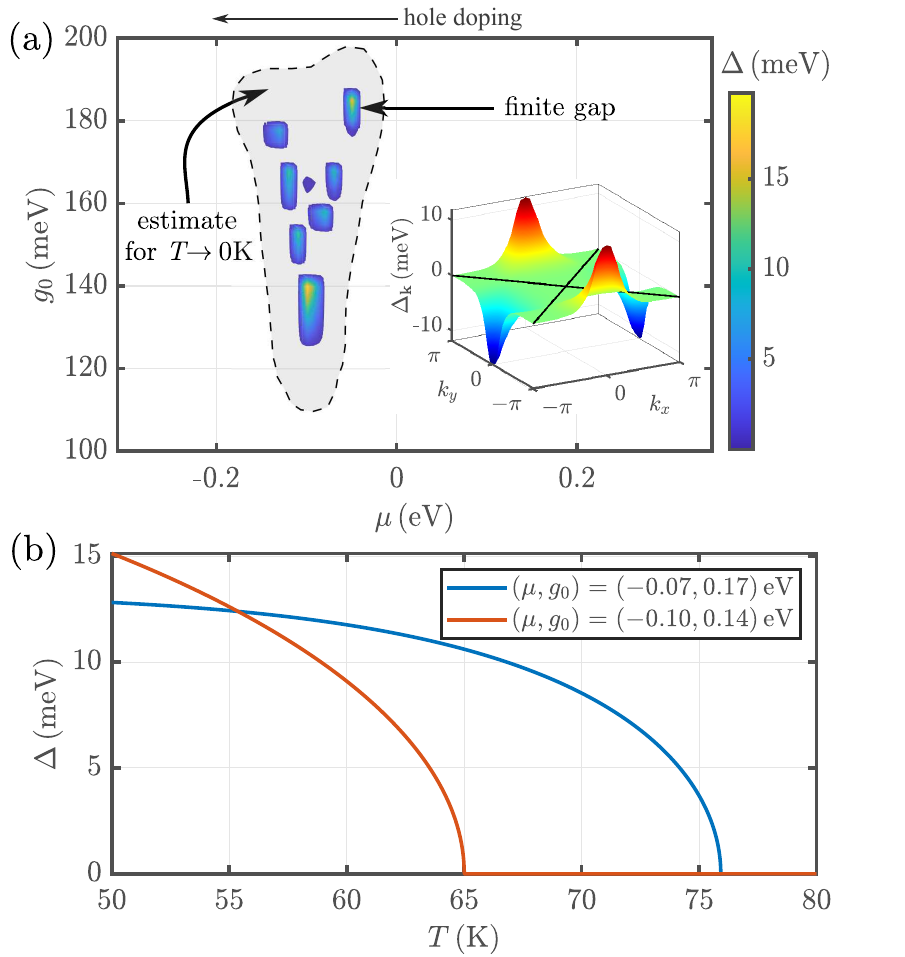}
	\caption{(a) Calculated  superconducting gap {$\Delta$} for the cuprate model as function of {\blue the initial} chemical potential $\mu$ and electron-phonon scattering strength $g_0$, obtained at $T=60\,\mathrm{K}$. The gray area depicts an estimate for finite superconducting solutions as temperature is decreased to zero. In the inset we show the {computed} momentum dependent zero-frequency result for the superconducting gap, {having $d_{x^2-y^2}$ symmetry,}  at $g_0=170\,\mathrm{meV}$ and $\mu=-0.07\,\mathrm{eV}$. (b) Temperature dependence of the gap $\Delta$, {computed} for two examples of $(\mu,g_0)$.}	\label{phasediag}
\end{figure}

Figure \ref{phasediag}(a) shows the result for the computed maximum superconducting gap $\Delta=\max_{\mathbf{k}}|\Delta_{\mathbf{k},m=0}|$ at $T=60\,\mathrm{K}$. Note that all pairs of $(\mu,g_0)$ where we find a finite gap magnitude lead {self-consistently} to a $d_{x^2-y^2}$ {gap} symmetry in momentum space (see inset in \ref{phasediag}(a)), partially with a nematic contribution \cite{Schrodi2021}. Interestingly, superconductivity is found only for a relatively small interval with respect to doping, and the results further depend heavily on the interaction strength. The temperature $T=60\,\mathrm{K}$ chosen here is relatively high, which is why we can expect that the phase space of available solutions for $\Delta>0$ grows {wider}  as $T$ decreases. Having performed {a} few selected calculations at smaller temperatures, we carved out an estimate of electron-phonon mediated $d$-wave superconductivity in the limit $T\rightarrow0\,\mathrm{K}$, {given by the gray shaded area in Fig.\ \ref{phasediag}(a).} 
An interesting feature of the gray shaded area is that we find it to be bounded even when the electron-phonon scattering strength $g_0$ becomes larger, i.e.,  counterintuitively, superconductivity disappears when $g_0$ increases. An explanation of this feature is given further below.

\begin{figure*}[t!]
	\centering
	\includegraphics[width=1\textwidth]{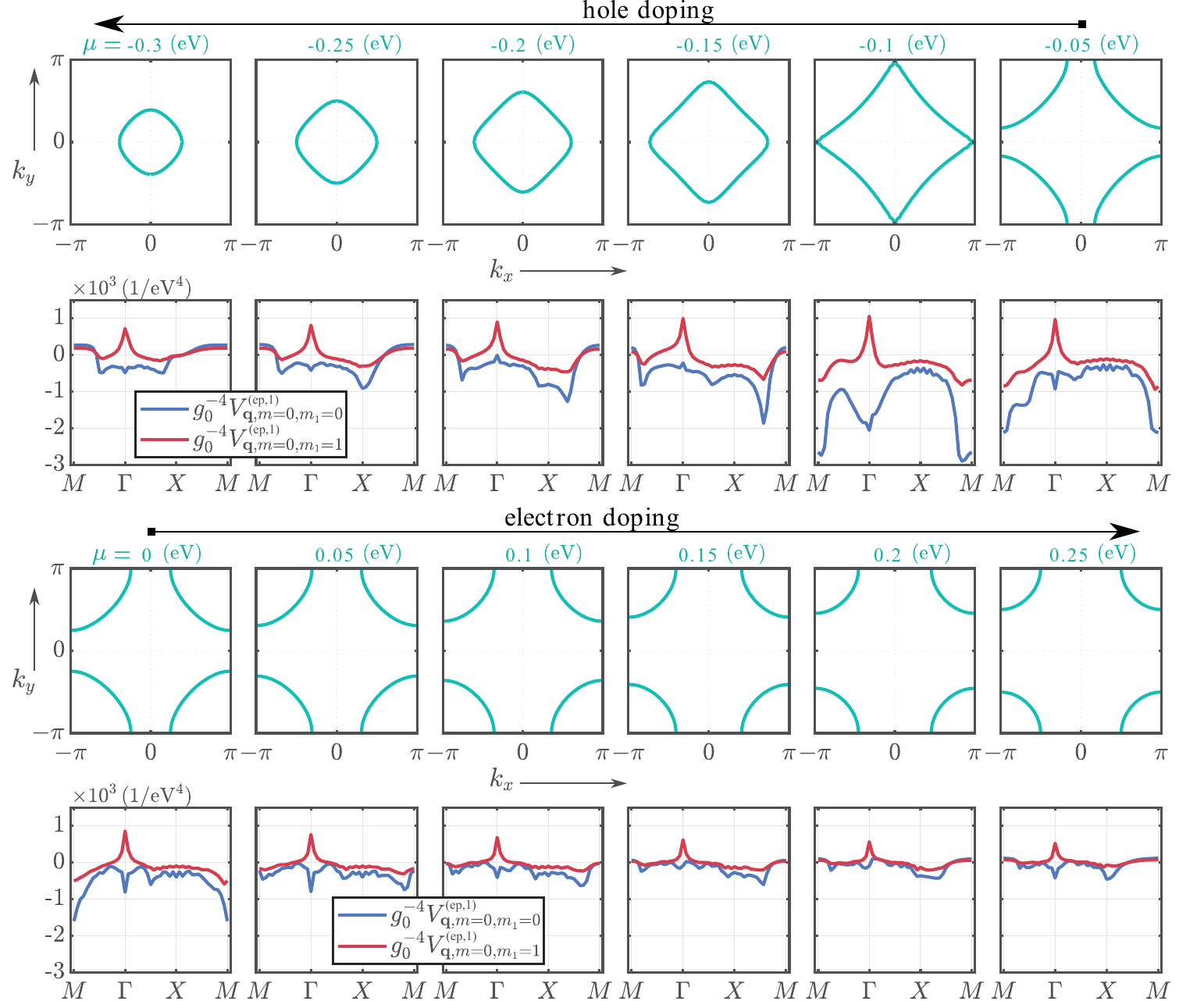}
	\caption{{Analysis of the} doping {dependence} of the electron-phonon interaction for $T=60\,\mathrm{K}$,  $\hbar \Omega=50\,\mathrm{meV}$ {\blue and $U=0$ eV}. The first and third rows show the evolution of the {\blue non-interacting} {cuprate} Fermi surface with {\blue initial} chemical potential $\mu$. Electron and hole doing is defined {\blue relative} 
	to {\blue the reference point} $\mu =0$. In the second and fourth row we show the associated interaction kernel $V^{\mathrm{(ep,1)}}_{{\bf q},m,m_1}$ from Eq.\,(\ref{kern1}), divided by $g_0^4$, along high-symmetry lines in the BZ, for frequency indices $(m,m_1)=(0,0)$ (blue) and $(m,m_1)=(0,1)$ (red). The high-symmetry points are labeled as $\Gamma =(0,0)$, $X=(\pi,0)$, and $M=(\pi,\pi)$. }	\label{dopingdep}
\end{figure*}

In Fig.\,\ref{phasediag}(b) we show the temperature dependence of $\Delta$ for two selected pairs $(\mu,g_0)$, see legend. It can be seen that the properties of the superconducting state can vary significantly, depending on doping level and coupling strength. For the blue curve we {obtain} $T_c\simeq76\,\mathrm{K}$ and $\Delta(T\rightarrow0\,\mathrm{K})/k_BT_c\simeq 2.02$, where the latter {\blue is} somewhat higher than the standard BCS value of 1.76. On the other hand, the red curve of Fig.\,\ref{phasediag}(b) leads to $T_c\simeq65\,\mathrm{K}$ and $\Delta(T\rightarrow0\,\mathrm{K})/k_BT_c\simeq4.25$, hence the effective coupling in the system is significantly increased, even though the bare coupling strength $g_0$ is smaller than for the previous example. Due to the fact that results from Fig.\,\ref{phasediag}(b) are obtained for different {\blue initial} chemical potentials, it deserves mentioning that a strong doping dependence of $\Delta(0)/k_BT_c$ in cuprate systems has both been found experimentally {\blue \cite{Tallon2022}} and theoretically for spin-fluctuation models \cite{Grabowski1996}.

To understand the results presented so far it is instructive to analyze the structure of the EPI.  We  consider the kernel function $V^{\mathrm{(ep,1)}}_{\mathbf{q},m,m_1} $ in more detail since its contribution is responsible for the unconventional gap symmetry in Eq.\,(\ref{phi}). 
We can isolate its doping dependence by calculating the corresponding  electron-phonon contributing term $V^{\mathrm{(ep,1)}}_{\mathbf{q},m,m_1}/g_0^4$ for different $\mu$ values.  In Fig.\,\ref{dopingdep} we show our results for the dominating lowest {Matsubara} frequency indices $(m,m_1)=(0,0)$ (blue) and $(m,m_1)=(0,1)$ (red) in the second and fourth rows, for different Fermi surface topologies as presented in the first and third rows. 
Note that the doping dependent Fermi surfaces are defined here with respect to $\mu$, relative to {\blue our chosen reference point} $\mu = 0$ eV, which  corresponds to the case of a typical {\blue hole-doped} cuprate Fermi surface. The latter is hole doped compared to the insulating copper oxide compound. {\blue Notably,} in our case, negative  $\mu$ corresponds thus to {\blue more hole doping and positive $\mu$  to less hole doping, or electron doping, relative to our reference point $\mu =0$.} {\blue We further observe that we obtain a steep decay of the superconducting gap for temperatures approaching $T_c$ which is possibly steeper than found in experiments \cite{Ren2012,Hashimoto2014}.}

From Fig.\,\ref{dopingdep} it is directly apparent that the red curve shows pronounced small-$\mathbf{q}$ contributions, i.e., a sharp peak around $\mathbf{q} =0$,  throughout the whole doping regime. The appearance of such a feature in second-order EPI has been predicted before \cite{Pietronero1995,Grimaldi1995}. Furthermore, the total magnitude of the interaction is biggest for the slightly hole-doped case {\blue ($\mu < 0$)}, still substantial for moderate electron doping, and becomes rather small for the extremely over- or under-doped system. For the blue curve the situation is similar in that the absolute values are largest close to $\mu =0$ and for small hole doping {\blue with respect to $\mu =0 $}. In contrast to the red curves, this function is purely negative throughout the entire BZ for many choices of $\mu$.

At this point it is important to stress that the curves shown in Fig.\,\ref{dopingdep} are not the overall interactions, but those results have to be multiplied by $g_0^4$ and added to the isotropic $V^{({\mathrm{ep}})}\propto g_0^2$. Since $V^{(\mathrm{ep})}>0$, the interplay between first and second order contributions can support a $d$-wave symmetric form of the superconducting order parameter if $V^{(\mathrm{ep})}+V^{(\mathrm{ep},1)}$ stays negative around the nesting wave vector $\sim (\pm\pi,\pm\pi)$ and positive for small $\mathbf{q}$ and thus depends sensitively on the presence of the nesting. As an example, for $\mu=-0.05\,\mathrm{eV}$ there exists a range of choices for $g_0$ that lead to an attractive small-$\mathbf{q}$ contribution and a repulsive long wavevector term, compare Fig.\,\ref{dopingdep}. The enhanced forward scattering peak of $(m,m_1)=(0,1)$ will further support the $d$-wave symmetry of the order parameter. If we consider  the strongly hole or electron doped regime, the first order contribution $V^{(\mathrm{ep})}$ to electron-phonon scattering in $V^{(\mathrm{ep})}+V^{(\mathrm{ep},1)}$ will however dominate for typical values of the scattering strength, hence a sign change of the total coupling is hard to achieve. Such interaction will then either support a conventional $s$-wave gap or self-consistently lead to a vanishing superconducting order parameter, which can happen for large $g_0$.

For an explicit {proof} that, for various doping levels, the experimentally observed $d$-wave symmetry of the superconducting gap is supported by the vertex-corrected electron-phonon coupling, we consider the purely momentum dependent interaction $V^{\mathrm{(ep,1)}}$ {\blue which appears in Eq.\ (\ref{phi}) for the superconducting order parameter.}  We define the form factor $f_{\mathbf{k}}^{(j)}$ that represents symmetry channel $j$. As a measure for the interaction strength $\lambda_j$ in this symmetry channel we calculate
\begin{equation}
\lambda_j(V^{\mathrm{(ep,1)}}_{\mathbf{q},m,m_1}) = \frac{1}{\mathcal{N}} \sum_{\mathbf{k},\mathbf{k}'} \frac{\delta(\xi_{\mathbf{k}})}{v_{\mathbf{k}}}  \frac{\delta(\xi_{\mathbf{k}'})}{v_{\mathbf{k}'}} f_{\mathbf{k}}^{(j)}V^{\mathrm{(ep,1)}}_{\mathbf{k}-\mathbf{k}',m,m_1}f_{\mathbf{k}'}^{(j)},
 \label{lambdaj}
\end{equation}
where $v_{\mathbf{k}}=|\nabla_{\mathbf{k}}\xi_{\mathbf{k}}|$ is the electron velocity, and $\delta(\xi_{\mathbf{k}})$ restricts the electronic degrees of freedom to the Fermi surface. The normalization is given by
\begin{equation}
\mathcal{N} = \sum_{\mathbf{k}} \frac{\delta(\xi_{\mathbf{k}})}{v^2_{\mathbf{k}}}   [f_{\mathbf{k}}^{(j)}]^2 ,
\end{equation}
and is, for simplicity, set to unity here, as we only want to analyze the doping dependence of $\lambda_j$.

\begin{figure}[h!]
	\centering
	\includegraphics[width=0.7\columnwidth]{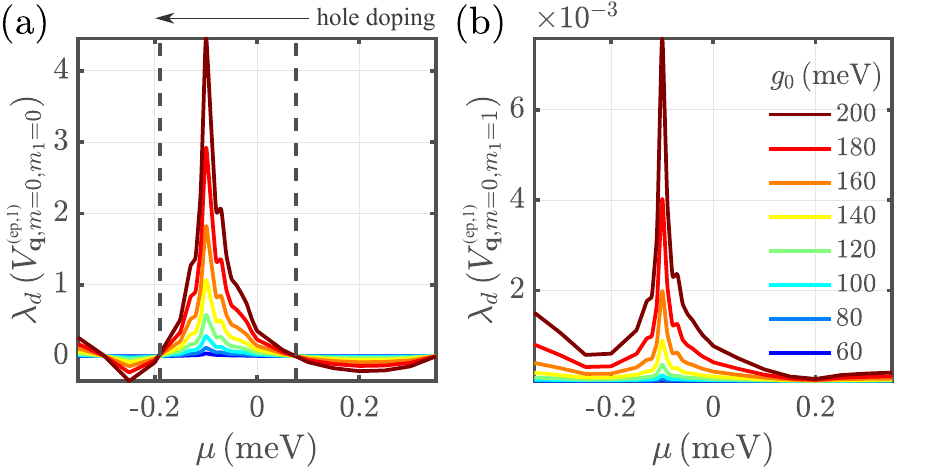}
	\caption{Projection of $V^{(\mathrm{ep,1})}_{\mathbf{q},m,m_1}$ onto the $d$-wave symmetry channel {\red of the gap} at $T=60\,\mathrm{K}$, {\red as a function of chemical potential $\mu$,} for (a) $m=m_1=0$ and (b) $m=0$, $m_1=1$. Different colors correspond to {\blue different} scattering strengths as listed in panel (b).}	\label{projection}
\end{figure}

Also, in this work we are only interested in the $d$-wave symmetry channel of the superconducting gap, which is the only experimentally relevant gap symmetry. It is characterized by the form factor $f_{\mathbf{k}}^{(d)}=\cos(k_x)-\cos(k_y)$. Hence, we compute the corresponding $\lambda_d$ for different choices of $g_0$ at $T=60\,\mathrm{K}$; the results are shown in Fig.\,\ref{projection}. As input for Eq.\,(\ref{lambdaj}) serves the interaction $V^{(\mathrm{ep,1})}_{\mathbf{q},m,m_1}$ for $(m,m_1)=(0,0)$  and $(m,m_1)=(0,1)$, with corresponding results for $\lambda_d$ shown in Figs.\,\ref{projection}(a) and (b), respectively. For both frequency index selections we find a pronounced peak of $\lambda_d$ in the slightly hole doped regime, with a maximum at around $\mu=-0.1\,\mathrm{eV}$. {Furthermore}, in Fig.\,\ref{projection}(a) we indicate the interval in which $\lambda_d >0 $ and thus a $d$-wave gap could appear by the dashed black lines. 

From Fig.\,\ref{projection} it becomes directly clear that the most important contribution to the interaction stems from frequency indices $m=m_1=0$, since the corresponding results due to $(m,m_1)=(0,1)$ are three orders of magnitude smaller. Therefore, we expect that the enhanced small-$\mathbf{q}$ couplings of Fig.\,\ref{dopingdep} play in fact a very minor role, which is different from predictions {made in} previous works \cite{Pietronero1995,Grimaldi1995}. Our results point toward the zero-frequency contribution to the interaction being mainly responsible for electron-phonon mediated $d$-wave superconductivity in the hole-doped system. 

Summarizing, our investigation of the doping dependence of the superconductivity  reveals that there is a realistic parameter regime $(\mu,g_0)$ where the vertex-corrected interaction mediates or supports an unconventional $d$-wave symmetry of the superconducting gap, which is  the gap symmetry observed in experiments \cite{Sobota2021,Chen1994,Shen1995,Kirtley1995,Shi2008,Ding1996}. Variation of the {\blue initial} chemical potential to simulate doping dependence uncovers that the doping level alters the shape of the Fermi surface and its nesting properties. This, in turn, strongly affects the EPI kernel, being negative around the nesting vector but positive (attractive) for small $\mathbf{q}$ vectors. Correspondingly, self-consistent $d$-wave superconductivity is only found in a relatively narrow doping interval around $\mu \approx  -0.1$ eV.
{\blue These findings are consistent with the well-known appearance of a superconducting dome around optimal doping in hole-doped cuprates \cite{Sobota2021}.}

\subsection{Influence of spin and charge fluctuations}
\label{SFs}

As a next step, we investigate the influence of spin and charge fluctuations in comparison to the electron-phonon coupling.
To do so, we treat the electron-phonon scattering strength $g_0$ and the on-site 
$U$ are free parameters in the following and compute their influence on the self-consistent superconductivity. {\blue We keep the 
 Einstein phonon frequency fixed at 50 meV and use for the chemical potential parameter $\mu = -0.07$ eV.} 

An extreme case would be to only consider spin and charge fluctuations, and no influence of EPI, i.e.\ choosing $g_0=0\,\mathrm{meV}$. For this setting it was not possible to stabilize a self-consistent solution {for the temperatures considered here, $T \ge 30$\,K,} regardless of the choice of $U$ and $T$ (minimum temperature tested was $T=30\,\mathrm{K}$). This result qualitatively agrees with recent studies on Fe-based superconductors \cite{Yamase2020,Schrodi2020_3} and we comment on it further below. 

{To check the consistency of the current approach with earlier work \cite{Lenck1994}, we tested smaller temperatures down to $T=6$\,K for the limiting case of vanishing $g_0$. It turns out that a finite superconducting gap with $d_{x^2-y^2}$ symmetry is obtained at temperatures smaller than 15\,K when choosing $U=0.35$\,eV. This choice puts the system in very close vicinity of a magnetic instability, which is reflected in an almost delta-peak-like shape of the SFs kernel. In the limit of $T \rightarrow 0$\,K we {computed} a gap magnitude of 5.4\,meV. These {results}, accompanied by unusually large values of the electron mass renomalization function, are in excellent agreement with previous work \cite{Lenck1994b}.} 

\begin{figure}[tb!]
	\centering
	\includegraphics[width=0.7\columnwidth]{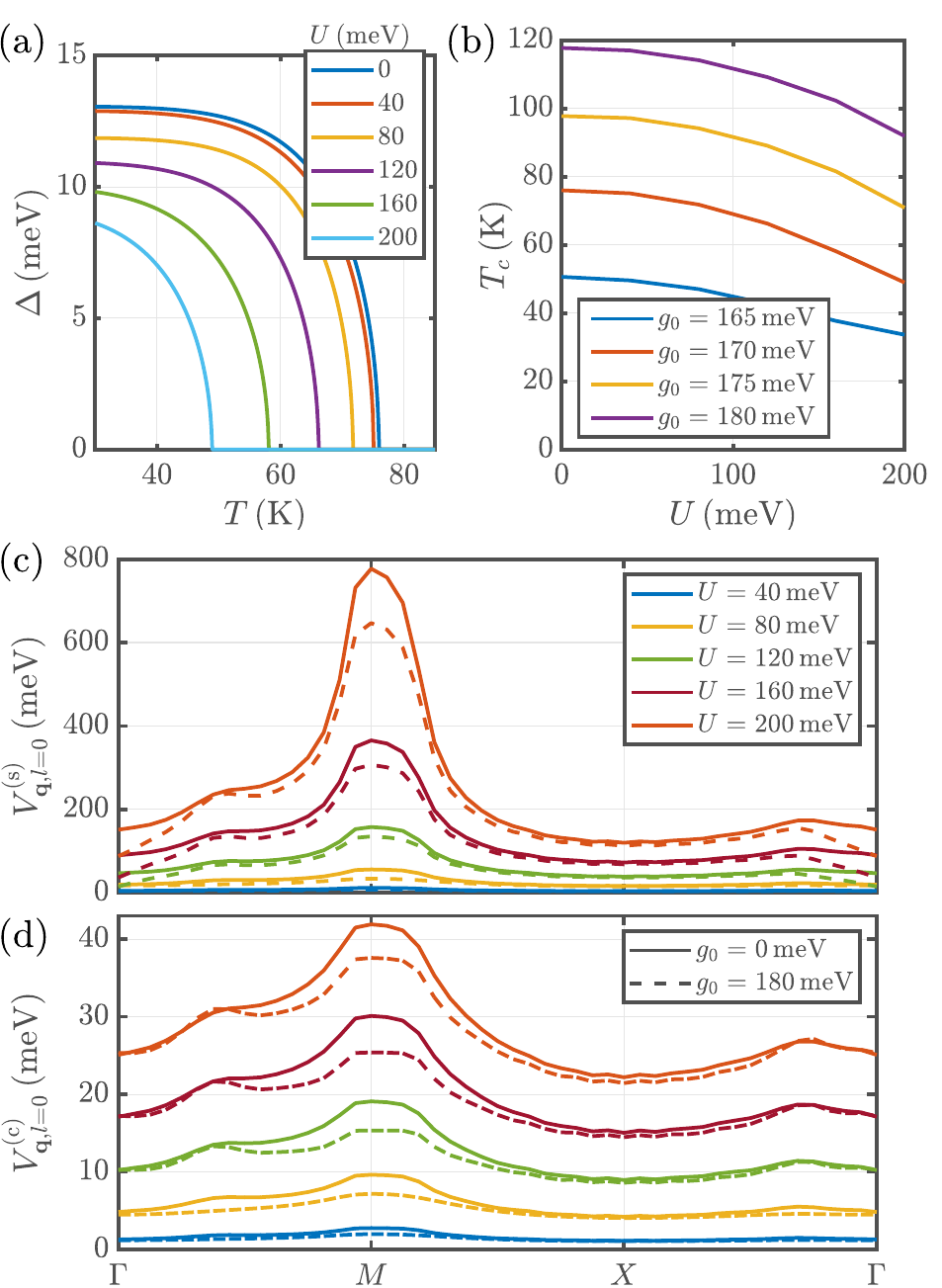}
	\caption{(a) Maximum superconducting gap as function of temperature, calculated for $g_0=170\,\mathrm{meV}$ and different values for $U$ as indicated in the legend. (b) Superconducting critical temperature $T_c$ against $U$ for varying choices of electron-phonon coupling constant $g_0$, see legend. Panels (c) and (d) show the zero-frequency component of the self-consistently obtained spin- and charge-fluctuations kernel, respectively, along high-symmetry lines of the BZ at $T=80\,\mathrm{K}$. Different colors correspond to varying choices of $U$, while solid and dashed lines represent $g_0=0\,\mathrm{meV}$ and $g_0=180\,\mathrm{meV}$, respectively.}
	\label{gaptc}
\end{figure}

To explore the influence of the 
$U$ parameter on the superconducting properties we show in Fig.\,\ref{gaptc}(a) the gap amplitude versus temperature for different choices of $U$ as indicated in the legend, and $g_0=170\,\mathrm{meV}$. As is directly apparent, the gap size and critical temperature decrease upon growing $U$. Following $T_c$ as function of $U$ in panel (b) of the same figure for different values of $g_0$ shows that this behavior is robust against changes in the electron-phonon coupling, and that therefore, 
{for the current level of approximation,} 
spin/charge fluctuations contribute destructively to the Cooper pairing.

To understand these findings we need to examine the self-consistently calculated spin and charge interactions. In Fig.\,\ref{gaptc}(c) and (d) we show $V_{\mathbf{q},l=0}^{(\mathrm{s})}$ and $V_{\mathbf{q},l=0}^{(\mathrm{c})}$, respectively. In both panels different values for $U$ are represented by the choice of colors, while solid and dashed lines correspond to $g_0=0\,\mathrm{meV}$ and $g_0=180\,\mathrm{meV}$, i.e., the absence or presence of EPI. We observe that the kernel from charge fluctuations is approximately one order of magnitude smaller than the spin-fluctuations part. Furthermore, as we increase the 
$U$ parameter, both interactions grow throughout the entire BZ, while $V_{\mathbf{q},l=0}^{(\mathrm{s})}$ tends to diverge around the nesting wave vector $\mathbf{q}\sim(\pi,\pi)$. 

It is important to note that the spin-fluctuations kernel enters repulsively in the equation for the superconducting order parameter, compare Eq.\,(\ref{phi}), which means that the interaction at exchange wave vector $\mathbf{q}\sim(\pi,\pi)$ (M) promotes a $d$-wave symmetry of the gap. However, in the absence of EPI the repulsive kernel around $\mathbf{q}  \sim(0,0) $ does not support this symmetry, which leads to phase oscillations {during the self-consistent iteration cycles,}
and, eventually, to $\Delta=0$. This behavior is analogous to that found in recent theoretical studies on Fe-based superconductors \cite{Yamase2020,Schrodi2020_3,Schrodi2020_4,Schrodi2021_3}. 

Conversely, in the presence of EPI ($g_0=  180$\,meV) we find a stable superconducting solution to the Eliashberg equations, which reduces the spin and charge kernels slightly throughout the entire BZ [see Figs.\,\ref{gaptc}(c) and (d)]. Most importantly, we see a significant decrease in $V_{\mathbf{q},l=0}^{(\mathrm{s})}$ around $\mathbf{q}=(0,0)$, hence the system {\blue self-consistently} adjusts the repulsive small-$\mathbf{q}$ contributions of the SFs, thereby promoting the Cooper pair formation. These adjustments of the interaction kernels underline that it is essential to achieve fully self-consistent solutions of the superconducting order parameter.

To summarize this part, we found for the considered model cuprate system that the EPI is responsible for the gap magnitude and the high critical temperature, while the SF part of the electronic interaction, in particular at small wave vectors, {\red is anti-pairing and} acts to reduce the superconductivity. At large wave vectors, on the other hand, both EPI and SF interactions cooperatively support the $d$-wave symmetry order parameter via repulsive 
coupling.

\section{Discussion and conclusions}

In this work we studied the effects of vertex-corrected electron-phonon interaction as well as of SFs and CFs on the superconductivity in a cuprate model system. We first investigated the influence of EPI scattering strength and doping level and then studied self-consistently the appearance of multichannel superconductivity, by treating vertex-corrected EPI, SFs and CFs together in one computational formalism. Our methodology opens a route for computationally studying anomalous 
non-adiabatic multichannel superconductors in a systematic way, which can be achieved via materials' specific choices of the phonon frequency, electron energy dispersion, and coupling strengths for both mediators of superconductivity. 

Our model system has roughly the correct characteristics of the family of compounds that we want to investigate, i.e., the {\blue hole}-doped cuprates with their prototypical Fermi surface.  In addition, the phonon frequency, EPI coupling strength $\lambda_m$, and electron-phonon scattering strength $g_0$ are in realistic ranges \cite{Lanzara2001,Iwasawa2008,Wang2021}. For the SF part of the interaction, a model similar to ours has been shown to be accurate for many characteristics of the interacting state \cite{Moriya2003}.

At this point it is appropriate to discuss the limitations of the here-developed computational approach. First, for numerical accessibility we neglected the effects of direct Coulomb repulsion and finite phonon self-energy.  The former is assumed here to be included in the electron energies $\xi_{\bf k}$. We expect that our results qualitatively hold even if both effects are {explicitly} included, yet a step toward a more complete picture  would be  the  inclusion  of  phonon  renormalization  effects and  the  Coulomb  repulsion. Second, we applied a non-interacting state approximation to the electron-phonon vertex correction, an approximation which we have shown to produce very accurate results compared to the full renormalized interaction for the here-studied cuprate system (see Fig.\,\ref{benchmark}). We note however that this might not be the case for other non-adiabatic superconductors.
Third,  while vertex-corrected Eliashberg theory that includes second-order scattering processes is expected to be valid for weakly non-adiabatic systems, higher order scattering processes might well play a role, especially for stronger non-adiabatic systems. 
{\red In the here-considered model system, the vertex-correction becomes larger than the isotropic lowest-order electron-phonon kernel at certain wave vectors, while being quite small in most of the BZ. This could signal that higher-order scattering processes could become relevant. It would thus be desirable to include in the future higher-order  Feynman diagrams for the electron-phonon interaction to assess self-consistently their influence on the superconductivity. Currently, this is however computationally out of the question.
In the strongly non-adiabatic case,} a competing charge density wave instability can occur \cite{Dee2023}, which is however not taken into account in our formalism. {\blue Monte Carlo simulations have recently investigated the competition of superconductivity and a charge-density instability, but considered only $s$-wave Cooper pairing \cite{Esterlis2018,Nosarzewski2021};  it would thus be interesting to self-consistently 
{\red treat} sign-changing $d$-wave pairing.}   
 Fourth, a renormalization of the SF spectrum due to direct magnon-phonon  coupling is not accounted for. Lastly, we mention that the SFs and CFs are treated in a single orbital model, but for a more realistic description a multi-orbital model might be required  \cite{Takimoto2004}. 
{\blue Including intra-orbital and inter-orbital on-site Hubbard-type interactions could possibly enlarge the phase space for obtaining non-zero selfconsistent gap solutions in the absence of electron-phonon coupling \cite{Schrodi2020_3}.} 
On a more general note, it deserves to be mentioned that the question about effects due to vertex corrections to the purely electronic interactions remains unanswered, which is of particular importance due to the absence of an analogue of Migdal's theorem for SFs \cite{Hertz1976}.

Our study of the doping dependence reveals that there is a realistic parameter regime with respect to the electron-phonon scattering strength, where the vertex-corrected interaction gives rise to an unconventional $d$-wave symmetry of the superconducting gap,  compatible 
 with experiments \cite{Chen1994,Shen1995,Kirtley1995,Shi2008}. It has been shown recently \cite{Schrodi2021} that nesting properties of the Fermi surface are instrumental for the sign change in the effective EPI which is a behavior that is very similar to the spin-fluctuations mechanism \cite{Scalapino2012}. Based on the Fermi surface topologies in Fig.\,\ref{dopingdep} we find the best nesting conditions for a slightly hole-doped system (relative to {\blue initial} $\mu =0$) which coincides with the largest coupling strength in the $d$-wave channel.  Furthermore, we find that superconductivity is obtained only in a relatively narrow doping range around the {\blue optimal}  hole-doped case, which correlates with the superconductivity dome in the {\blue hole-doped region of the} temperature-doping phase diagram that is prototypical for the cuprates \cite{Sobota2021}.

Our results from Fig.\,\ref{phasediag} further indicate that EPI alone can lead to realistic values of the gap magnitude and $T_c$, for certain reasonable choices of the coupling strength. {\red Conversely, including only the electronic SFs and CFs interactions leads to particularly low values of $T_c$ and the gap size.} Considering that such results are found from a model study, we speculate that some properties of the superconducting state in the actual members of the cuprate family may differ due to varying phonon frequencies, details of the Fermi surface topology, and electron-phonon coupling strengths. 

Applying our computational methodology to include both EPI and SFs/CFs as possible mediators of multichannel superconductivity \cite{Schrodi2020_4,Schrodi2021,Schrodi2021_3}, we have found that the EPI is mainly responsible for the gap magnitude and high critical transition temperature, while the SF part of the electronic interaction overall suppresses superconductivity. Both interactions however cooperate to support the $d$-wave symmetry order parameter via repulsive 
coupling {at large wave-vectors}. We found that repulsive small momentum wave-vector contributions from the SF kernel are responsible for the {\blue here-}reported suppression of $T_c$. Our self-consistent simulations showed however that the influence of repulsive SF interaction at  small-$\mathbf{q}$ in the system can  become reduced due to attractive EPI.  Since a similar behavior has been encountered in Fe-based compounds \cite{Yamase2020,Schrodi2020_4}, it might be a generic effect in high-temperature superconductors.

Lastly, we stress that out focus is not to fully explain high-temperature superconductivity in cuprates, but rather to clarify the competition between pairing mediators and the origin of its remarkable doping dependence on the basis of self-consistent vertex-corrected Eliashberg theory. We cannot exclude the possibility that our results might change when a more complete theory of self-consistent calculations of the superconducting state will be employed. Within the current approach, we find  that vertex-corrected electron-phonon coupling supports superconductivity with an unconventional gap symmetry only in a relatively small doping interval, consistent with the superconductivity dome in the {\blue hole-doped cuprate} phase diagram.
We further find that both electron-phonon and spin-fluctuation mechanisms support unconventional $d$-wave symmetry, yet the EPI mainly mediates the pairing and is responsible for the high $T_c$, whereas SFs are found to have a suppressing influence on both the gap and $T_c$.

\ack
This work has been supported by the Swedish Research Council (VR) and the Knut and Alice Wallenberg Foundation (grants No.\ 2022.0079 {\blue and 2023.0336}).	
{The Eliashberg-theory calculations were enabled by resources provided by the National Academic Infrastructure for Supercomputing in Sweden  (NAISS) at NSC Link{\"o}ping, partially funded by VR through Grant Agreement No. 2022-06725.}

\section*{References}
\providecommand{\newblock}{}

\end{document}